\documentclass[conference]{IEEEtran}
\usepackage{amsmath}
\usepackage{graphicx}
\usepackage{multirow}
\usepackage{nomencl}
\usepackage{cite}
\usepackage{graphicx,dblfloatfix}
\usepackage{epstopdf}
\usepackage{algorithm}
\usepackage{algorithmic}
\usepackage{balance}
\usepackage{comment}
\usepackage[keeplastbox]{flushend}
\usepackage{mathtools}
\usepackage{amsthm}
\usepackage{amssymb}
\usepackage{color}
\usepackage{float}

\newcommand{\hh}{{\mathcal{H}_2}}
\newcommand{\hi}{{\mathcal{H}_{\infty}}}
\newcommand{\ab}{{\big \vert \! \big \vert}}

\newtheorem{proposition}{Proposition}

\newtheorem{definition}{Definition}

\IEEEoverridecommandlockouts
\usepackage{eso-pic}
\newcommand\AtPageUpperMyright[1]{\AtPageUpperLeft{%
		\put(\LenToUnit{0cm},\LenToUnit{-1cm}){%
			\parbox{0.54\textwidth}{\raggedleft\fontsize{9}{11}\selectfont #1}}%
}}%
\newcommand{\conf}[1]{%
	\AddToShipoutPictureBG*{%
		\AtPageUpperMyright{#1}
	}
}    
\conf{American Control Conference (ACC), Milwaukee, WI, 2018.}
\begin{document}
	\renewcommand*\footnoterule{}
	\title{Optimal $\hh$ Decentralized Control of Cone Causal Spatially Invariant Systems}
	\author{M. Ehsan Raoufat,~\IEEEmembership{Student Member,~IEEE,}
		    Seddik M. Djouadi,~\IEEEmembership{Member,~IEEE,}
		\thanks{M. Ehsan Raoufat and Seddik M. Djouadi are with the Min H. Kao Department of Electrical Engineering and Computer Science, University of Tennessee Knoxville, TN 37996 USA (e-mail: mraoufat@utk.edu).}}
	\maketitle
\begin{abstract}
	\boldmath
	This paper presents an explicit solution to decentralized control of a class of spatially invariant systems. The problem of optimal $\hh$ decentralized control for cone causal systems is formulated. Using Parseval's identity, the optimal $\hh$ decentralized control problem is transformed into an infinite number of model matching problems with a specific structure that can be solved efficiently. In addition, the closed-form expression (explicit formula) of the decentralized controller is derived for the first time. In particular, it is shown that the optimal decentralized controller is given by a specific positive feedback scheme. A constructive procedure to obtain the state-space representation of the decentralized controller is provided. A numerical example is given and compared with previous works which demonstrate the effectiveness of the proposed method.  
\\
\end{abstract}
\renewcommand\IEEEkeywordsname{Index Terms}

\section{Introduction}
Decentralized control problems, different from classical centralized ones, have received considerable attention in recent years. These problems arise when a system consists of several decision makers (DMs) in which their actions are based on decentralized information structure \cite{mahajan2012information}. The term decentralized used in this paper is a general term where decisions are based on a subset of the total information available about the system. The information structure has a direct impact on the scalability and tractability of computing optimal decentralized controllers \cite{mahajan2008identifying}. Over the past few years, different decentralized information structure has been analyzed in detail \cite{mahajan2012information, mahajan2008identifying, mahajan2009optimal, nayyar2011optimal}.  \\ \indent
In practice, most complex systems consist of interconnections of many subsystems. Each subsystem may interact with its neighbors and include several sensor and actuator arrays. Examples of such systems include coordination of large-scale power systems\cite{Ehsan} and flight formation \cite{wolfe1996decentralized}. In some cases, lumped approximations of PDEs can also be used for modeling and control of identical interconnected systems such as distributed heating/sensing \cite{bamieh1999optimal}, and large vehicle platoons \cite{barooah2007control}. For these spatially distributed systems, centralized strategies are computationally expensive and might be impractical in terms of hardware limitations such as communication speed. Hence, decentralized control strategies are more desirable. 
\\ \indent
Decentralized control problems were first studied by Radner as a team decision problem \cite{radner1962team}. Major difficulties that arise in these problems are from the information patterns. Early work by Witsenhausen \cite{witsenhausen1968counterexample} demonstrated the computational difficulties associated with team decision making even in a simple two-player process. In \cite{ho1972team}, it has been shown that for a unit-delay information sharing pattern, the optimal controller is linear. However, decentralized control problems with multi-step delayed information sharing patterns generally do not have this property. \\ \indent
In this paper, we study the decentralized control of spatially invariant systems that are made up of infinite numbers of identical subsystems and are functions of both temporal and spatial variables. Spatial invariance means that the distributed system is symmetric in the spatial structure and the dynamics do not vary as we shift along some spatial coordinates. In \cite{bamieh1999optimal}, a framework for spatially invariant systems with distributed sensing and actuation has been proposed and optimal control problems such as LQR, $\hh$ and $\hi$ has been studied in a centralized fashion. In \cite{bamieh2002distributed}, the authors claimed that for spatially invariant systems, dependence of the optimal controller on information decays exponentially in space and the controller have some degree of decentralization.  \\ \indent
Decentralized control problems can also be reformulated in a model matching framework using the Youla parametrization. For general systems, this nonlinear mapping from the controller to the Youla parameter removes the convexity of constraint sets (e.g. decentralized structure) \cite{rotkowitz2010parametrization}. However, a large class of systems called quadratic invariance have been introduced in \cite{rotkowitz2006characterization}, under which the constraint set is invariant under the above transformation. Different constraint classes such as distributed control with delay, decentralized control, and sparsity constraints have been considered in the literature, and methods such as vectorization \cite{rotkowitz2006characterization} or optimization based techniques \cite{lamperski2013output} have been used for computation. However, no explicit solution has been provided and due to high computation requirements and numerical issues, vectorization approach is limited to systems with a small number of states \cite{kim2015explicit}.\\ \indent
Distributed controller design problem for classes of spatially invariant systems with limited communications can also be cast as a convex problem using Youla parametrization. This class includes spatially invariant systems with additional cone causal property where information propagates with a time delay equal to their spatial distance \cite{voulgaris2003optimal}. A more general class than cone causality, termed as funnel causality where the propagation speeds in the controller are at least as fast as that of the plant was introduced in \cite{bamieh2005convex}. It is important to note that decentralized control with structures such as cone or funnel causality yields a convex problem. However, these problems are in general infinite dimensional and finding the explicit solutions or developing an efficient procedure for solving them are still open, and are the subject of intense research. 
\\ \indent
In our previous work \cite{djouadi2014duality}, the optimal centralized control problem for this type of spatially invariant systems have successfully been posed as a distance minimization in a general $L_{\infty}$ space, from a vector function to a subspace with a mixed $L_{\infty}$ and $\mathcal{H}_{\infty}$ space structure. In \cite{djouadi2015operator}, we have formulated the Banach space duality structure of the problem in terms of tensor product spaces.
In \cite{djouadi2015distributed}, the optimal centralized and decentralized $\hh$ control problem is formulated using an orthogonal projection from a tensor Hilbert space of $L_2$ and $\hh$ onto a particular subspace.  However, further extensions are needed to solve the problem explicitly and calculate the transfer function or state-space characterization. 
\\ \indent
Motivated by the concern outlined above, in this paper, the optimal $\hh$ decentralized control problem for a class of spatially invariant system is considered. By building on our previous results, the decentralized problem for cone causal systems is derived. Using Parseval's identity, the optimal $\hh$ decentralized control problem is transformed into an infinite number of model matching problems with a specific structure that can be solved efficiently. In addition, the closed-form expression (explicit formula) of the optimal decentralized controller is derived which was previously unknown. A constructive procedure to obtain the state-space representation of the decentralized controller is also provided. A numerical example is given with comparison with previous works demonstrates the effectiveness of the proposed method. \\ \indent
The paper is organized as follows. In Section II, mathematical preliminaries and state-space representation of discrete cone causal systems are presented. In Section III, we demonstrate that the optimal decentralized controller can be designed based on an infinite number of model matching problems which can be solved efficiently. This is followed in Section IV by an example illustrating the validity of the results. Finally, some concluding remarks are drawn in Section V.

\section{Discrete Cone Causal Systems}
\begin{figure}[!t]
	\centering \vspace{-0.5cm}
	\includegraphics[width=3.1in]{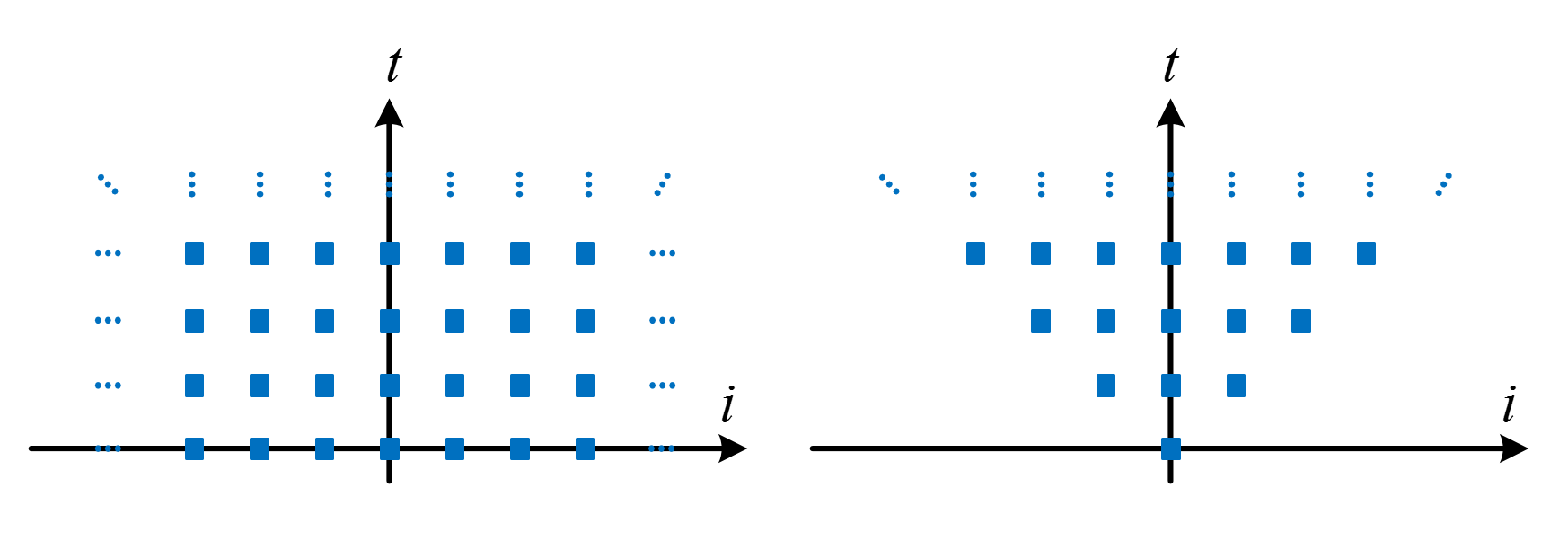} \vspace{-0.5cm}
	\caption{Support of the spatio-temporal impulse response of a centralized (left) and a cone causal (right) system.}
	\label{fig:1}
\end{figure}

The framework considered in this paper for discrete cone casual systems was first introduced in \cite{voulgaris2003optimal}. The spatially invariant system $G$ with inputs $u(i,t)$ and outputs $y(i,t)$ has the following form
\begin{IEEEeqnarray}{c}
	y(i,t)=\sum_{j=-\infty}^{\infty}{\sum_{\tau=-\infty}^{\infty}{\hat{g}(i-j,t-\tau)u(j, \tau)}}  \\
    \hat{g}(i,t)=0, \quad \forall t<0\;\; ({\rm due\;to\;temporal\; causality}) \nonumber
\end{IEEEeqnarray}
where $i$ is discrete space, $t$ is discrete time, and $\hat{g}(i,t)$ represents the spatio-temporal impulse response of $G$ and has temporal causality. Using the $\lambda$-transform $g(i,\lambda)=\sum_{t=0}^{\infty}{\hat{g}(i,t)\lambda^t}$, spatio-temporal transfer function $G$ is given by $G(z,\lambda) :=\sum_{i=-\infty}^{\infty}{g(i,\lambda)z^i}$ where $z$ denotes the two-sided spatial transform variable, $\lambda$ denotes the one-sided temporal transform variable and input-output relation is as follow
\begin{IEEEeqnarray}{c}
	Y(z,\lambda)=G(z,\lambda)U(z,\lambda)
\end{IEEEeqnarray}
where $Y(z,\lambda)$ is the transform of $y(i,t)$ and $U(z,\lambda)$ is the transform of $u(i,t)$.
Particular structure of interest is the case where the spatio-temporal impulse response of the system $\hat{g}(i,t)$ has the cone causal structure as shown in Fig.~\ref{fig:1}.
\begin{figure}[!t]
	\centering \vspace{-0.4cm}
	\includegraphics[width=3.1in]{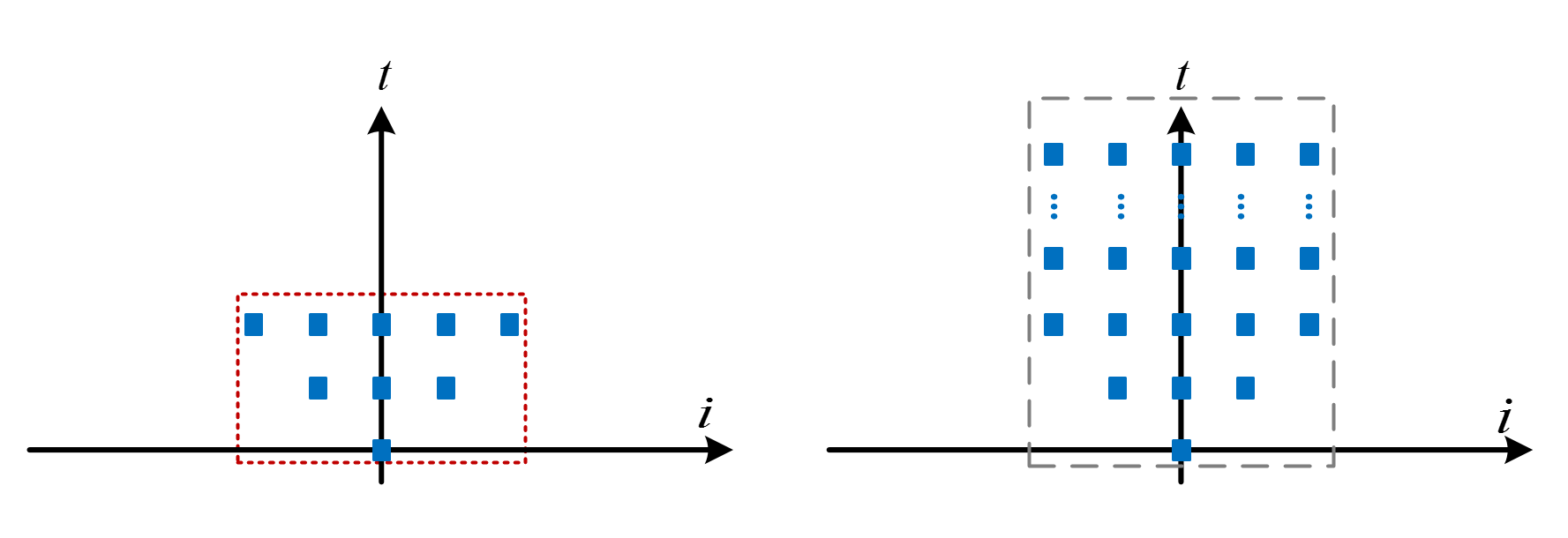} \vspace{-0.5cm}
	\caption{Support of a finite approximation of a cone causal system using equation (\ref{Eq:G_cone}) in the right and equation (\ref{Eq:G_cone_other}) in the left.}
	\label{fig:2} \vspace{-0.2cm}
\end{figure}
\begin{definition}
	A discrete linear system $y=Gu$ is called cone causal if it has the following form \cite{voulgaris2003optimal}
	\begin{eqnarray}
	G(z, \lambda) &=& \sum\limits_{i=-\infty}^{\infty} g_i(\lambda)z^i    \label{Eq:G_cone} \\
	g_i(\lambda) &=& \lambda^{|i|}\tilde{g}_i(\lambda)  \nonumber
	\end{eqnarray}
	 where the transfer function $\tilde{g}_i(\lambda)$ corresponds to temporally causal systems.
\end{definition}
The interpretation of this property is that the input $u_k$ to the $k$th system $g_k$ affects the output $y_m$ of the $m$th system $g_m$, which is $|k-m|$ spatial location away with a delay of $|k-m|$ time steps \cite{voulgaris2003optimal}. This type of cone causal systems can also be written in the following form
\begin{IEEEeqnarray}{rlc}
	G(z,\lambda) &=& \sum_{k=0}^{\infty}g_k(z) \lambda^k  \label{Eq:G_cone_other}\\
	g_k(z) &=& \sum_{n=-k}^{k}g_{n,k} z^n  \nonumber
\end{IEEEeqnarray}
For an infinite number of terms, the above two systems are equivalent. However, we usually use a finite number of terms in the calculations, thus the first definition is more general as shown in Fig. \ref{fig:2} for one example. \\ \indent
In general, the transfer function $G(z,\lambda)$ can be seen as a multiplication operator on $\mathcal{L}_2(\mathbf{T} ,\bar{\mathcal{D}})$ where $\mathbf{T}$ is the unit circle and $\bar{\mathcal{D}}(\mathcal{D})$ is the closed (open) unit disc of the complex domain $\Bbb{C}$. Assume that $G(z,\lambda)$ is stable, then we have \cite{djouadi2015distributed}
\begin{IEEEeqnarray}{rlc}
	G(z,\lambda): \; \mathcal{L}_2(\mathbf{T},\bar{\mathcal{D}}) \; &\longrightarrow& \quad \mathcal{L}_2(\mathbf{T},\bar{\mathcal{D}}) \\
	u \; &\longrightarrow& \; Gu =G(e^{i \theta},\lambda)u(e^{i \theta},\lambda) \nonumber
\end{IEEEeqnarray}
where $0 \le \theta < 2\pi$, and $| \lambda | \le 1$. 
From $ H^p $-theory \cite{duren2000theory} asserts that if $ f\in \hh$, then $ f(e^{jw}) \in L_2 $, that is, $H_2$ may be viewed as a closed subspace of $L_2$. Letting $ \hh^{\perp} $ be the orthogonal complement in $ L_2 $, then we have
\begin{IEEEeqnarray}{rlc}
	L_2 = \hh \oplus \hh^{\perp},  \label{cdc08}
\end{IEEEeqnarray}
which means that every $ f \in L_2 $ can be written uniquely as $f = f_1 + f_2$ with $ f_1 \in H_2$ and $ f_2 \in \hh^{\perp} $. \\ \indent
The $\ell_2$-norm of the original system can be defined as
\begin{equation}
	\|G\|_2 = \left(\sum_{i=-\infty}^{\infty} \sum_{t=0}^{\infty} |\hat{g}(t,i)|^2 \right)^{\frac{1}{2}}      \label{cdc01}
\end{equation}
and the $\hh$-norm of its transform $G(z,\lambda)$ is given by
\begin{equation}
	\| G \|_{\hh}=\frac{1}{2\pi} \left[ \int_{\theta \in [0,2\pi]} \int_{w\in[0,2\pi]} \bigl|G (e^{i\theta}, e^{iw} ) \bigl|^2 dw d\theta \right]^{\frac{1}{2}}  	\label{cdc02}
\end{equation}
where the isometry $ \| G \|_2 = \| G\|_{\hh} $ holds. Before solving the optimal $\hh$ decentralized control problem for cone causal systems, it is important to review the basics of state-space representation of this class of systems.
\begin{definition}
	Consider the system $G$ with state-space representation
	\begin{IEEEeqnarray}{c}
		G=\left[ \begin{array}{c|c}
			A(z) & B \\ \hline
			C(z) & D
		\end{array} \right]	 = D + \lambda C(z)\big(I-\lambda A(z)\big)^{-1}B
	\end{IEEEeqnarray}
	The set of $\ell$-causal system refers to the system where $B$ and $D$ are independent of $z$ and matrices $A(z)$ and $C(z)$ are of the following forms \cite{fardad2011design}
	\begin{IEEEeqnarray}{c}
		A(z)=A_{-1}z^{-1} + A_{0} + A_{1}z^{1} \\
		C(z)=C_{-1}z^{-1} + C_{0} + C_{1}z^{1}
	\end{IEEEeqnarray}
where $A_n$ and $C_n$ are independent of $z$ and the dimension of the matrix $A$ denotes the temporal order of the system.
\end{definition}
The set of $\ell$-causal systems is equal to the set of cone causal systems. Note that this set is closed under addition, composition, and inversion of systems \cite{fardad2011design}. Thus, it is closed under feedback and linear fractional (Youla) transformations. For complex systems, the state space representation of the controller can be obtained by realizing each element of the transfer function by performing basic sum, product, and inverse operations. Suppose that $G_1$ and $G_2$ are two subsystems with the following state-space representation
\begin{IEEEeqnarray}{c}
	G_1=\left[ \begin{array}{c|c}  A_1(z) & B_1 \\ \hline  C_1(z) & D_1  \end{array} \right]	,  \quad
	G_2=\left[ \begin{array}{c|c}  A_2(z) & B_2 \\ \hline  C_2(z) & D_2  \end{array} \right]
\end{IEEEeqnarray}
The following operations are useful to build the state-space model of the transfer function \cite{zhou1998essentials}
\begin{IEEEeqnarray}{rlc}
	G^{-1}_1 &= \left[ \begin{array}{c|c}
		A_1(z)-B_1 D^{-1}_1 C_1(z) & -B_1 D^{-1}_1  \\ \hline
		D^{-1}_1 C_1(z) & D^{-1}_1
	\end{array} \right] \label{SS1}\\
	G_1 + G_2 &= \left[ \begin{array}{cc|c}
	A_1(z) 	& 0    & B_1 			\\
	0 		& A_2(z)  & B_2   \\ \hline
	C_1(z) 	& C_2(z)  & D_1+D_2
	\end{array} \right] \label{SS2}\\
	G_1 G_2 &= \left[ \begin{array}{cc|c}
	A_1(z) 	& B_1 C_2(z)  &  B_1 D_2		\\
	0 		& A_2(z)   	&  B_2   \\ \hline
	C_1(z) 	& D_1 C_2(z)   &  D_1 D_2
	\end{array} \label{SS3}\right]
\end{IEEEeqnarray}
In the next section, the decentralized problem is verbalized in detail and an explicit solution is presented.
\begin{figure}[!t]
	\centering \vspace{-0.4cm}
	\includegraphics[width=2.5in]{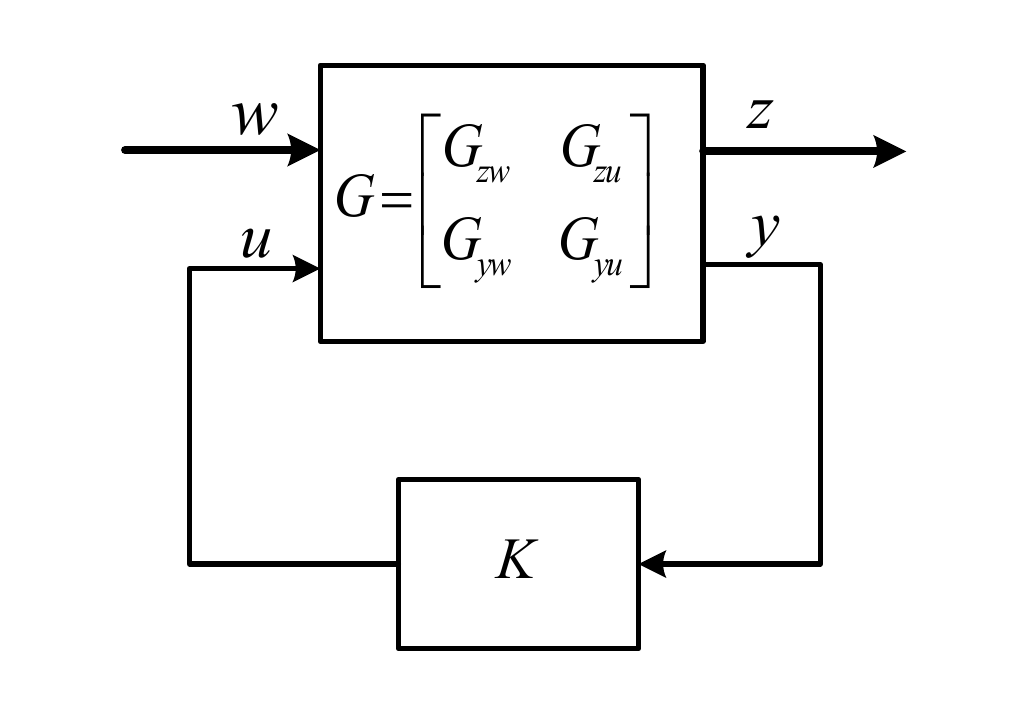} \vspace{-0.4cm}
	\caption{General control configuration.}  \vspace{-0.4cm}
	\label{fig:Dis_Attenuation}
\end{figure}

\section{Design of Optimal $\hh$ Decentralized Controller}
\begin{figure}[!t]
	\centering
	\includegraphics[width=2.6in]{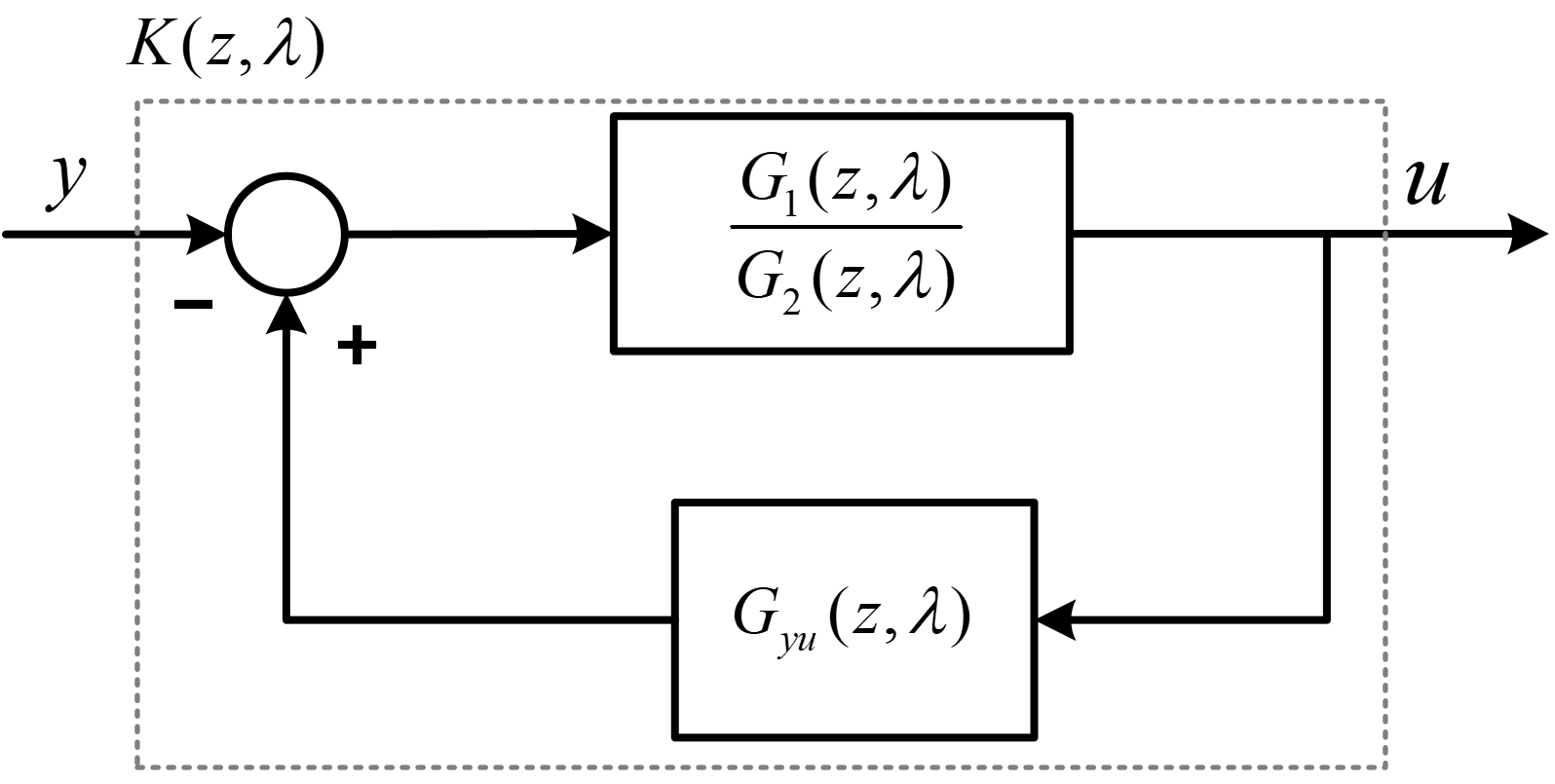}
	\caption{Positive feedback block diagram realization of $K$.} \vspace{-0.4cm}
	\label{fig:3}
\end{figure}

Our goal in this paper is to design the optimal $\hh$ decentralized controllers for general disturbance attenuation problem as shown Fig. \ref{fig:Dis_Attenuation}. The open loop system is denoted by $G$, the controller by $K$, the performance outputs by $z$, the measurements by $y$, the input control signals by $u$ and the external disturbances by $w$. The closed loop disturbance response from $w$ to $z$ is given by
\begin{eqnarray}
	T_{zw} = G_{zw} + G_{zu}K(I-G_{yu}K)^{-1}G_{yw}
\end{eqnarray}
where the stable spatio-temporal controller $K$ (internally) stabilizes and minimizes the $\hh$ norm of the disturbance transfer function $T_{zw}$. The particular structure of interest is when the spatio-temporal transfer function $G_{yu}$ yields the following cone causal form (as defined in Section II):
\begin{eqnarray}
	G_{yu}(z, \lambda) &=& \sum\limits_{i=-\infty}^{\infty} g_i(\lambda)z^i    \label{Eq:Gyu} \\
	g_i(\lambda) &=& \lambda^{|i|}\tilde{g}_i(\lambda)  \nonumber
\end{eqnarray}
The optimal decentralized controllers $K$ have the same structure as $G_{yu}$ \cite{voulgaris2003optimal,vidyasagar2011control}, that is, 
\begin{eqnarray}
	K(z,\lambda) &=& \sum\limits_{i=-\infty}^{\infty} k_i(\lambda)z^i  \label{Eq:K} \\
	k_i(\lambda) &=& \lambda^{|i|} \tilde{k}_i(\lambda)    \nonumber
\end{eqnarray}
which means that the measurements of the $j$th location will be available at the $i$th system after $|j-i|$ time steps delay. \\ \indent
The following proposition asserts that the decentralized constraints on the controller  $K$ can be enforced on the Youla parameter $Q$ using similar convex constraints.
\begin{proposition} \label{propo1}
	For open loop stable systems, all stabilizing controllers $K$ with the structure (\ref{Eq:K}) are given by \cite{voulgaris2003optimal,vidyasagar2011control}
	\begin{eqnarray}
		K = -Q(I-G_{yu}Q)^{-1}   \label{cdc28}
	\end{eqnarray}
	with $Q$ given by
	\begin{eqnarray}
		Q(z,\lambda) &=& \sum\limits_{i=-\infty}^{\infty} q_i(\lambda) z^i    \label{Eq:Q} \\
		q_i(\lambda) &=& \lambda^{|i|}\tilde{q}_i(\lambda)   \nonumber
	\end{eqnarray}
	where $\tilde{q}_i (\lambda)$ is stable.
\end{proposition}
Using the Youla parameterization, the disturbance transfer function can be recast as $T_{zw} = T_1 - T_2 Q$. The decentralized $\hh$ optimal control problem can then be written as
\begin{eqnarray}
	J &:=& \inf_{K_{stabilizing} \; {\rm s.t.} (\ref{Eq:K}) \; {\rm hold}} \| T_{zw}\|_{\hh}             \label{Eq:J} \nonumber \\
			  &=& \inf_{Q \, {\rm stable} \; {\rm s.t.} (\ref{Eq:Q}) \; {\rm hold}} \|T_1-T_2 Q \|_{\hh}
\end{eqnarray}
The inner-outer factorization of $T_2$ defined as
\begin{eqnarray}
	T_2(e^{j \theta},\lambda) = T_{2in}(e^{j \theta},\lambda) T_{2out}(e^{j \theta},\lambda)
\end{eqnarray}
where inner function $T_{2in}$ is isometry and outer function $T_{2out}$ is causally invertible. Therefore, (\ref{Eq:J}) reduces to
\begin{eqnarray}
	J &=& \inf_{Q \, {\rm stable} \; {\rm s.t.} (\ref{Eq:Q}) \; {\rm hold} } \| T_{2in}^\ast T_1-T_{2out}Q \|_{L_2}  \label{J_simp}	\quad
\end{eqnarray}
Since $ \{ z^i \}_{i=-\infty}^\infty $ is an orthogonal basis of $L_2$, $ T_{2in}^\ast T_1 $ can be written as:
\begin{eqnarray}
	T_{2in}^\ast (z,\lambda) T_1 (z,\lambda) = \sum\limits_{i=-\infty}^{\infty} \tilde{T}_i(\lambda)z^i  \label{T_i}
\end{eqnarray}
where $\tilde{T}_i \, (\lambda) \in L^2$.
The outer function $T_{2out}$ also admits the same cone structure
\begin{eqnarray}
	T_{2out}(z,\lambda) &=& \sum_{i = - \infty}^{\infty}v_i(\lambda) z^i      \label{Eq:Tout} \\
	v_i(\lambda) &=& \lambda^{\vert i \vert} \tilde{v}_i(\lambda)           \nonumber
\end{eqnarray}		
and $\tilde{v}_i(\lambda)$ is stable. Therefore, $T_{2out}Q$ have the following structure
\begin{IEEEeqnarray}{c}
	T_{2out}(z,\lambda)Q(z,\lambda) = \sum_{i = - \infty}^{\infty} \eta_i(\lambda) z^i \label{T2outQ}
\end{IEEEeqnarray}
where $\eta_i(\lambda)$ can be written as
\begin{eqnarray}
	\eta_i (\lambda) &=& \sum_{j = - \infty}^{\infty} \lambda^{\vert j \vert} \tilde{q}_j(\lambda)	\lambda^{\vert i-j \vert} \tilde{v}_{i-j}(\lambda)  \quad \nonumber\\
	&=& \sum_{j = - \infty}^{\infty} \lambda^{\vert j \vert + \vert i-j \vert} \tilde{q}_j(\lambda) \tilde{v}_{i-j}(\lambda)
\end{eqnarray}
From triangle inequality, we have
\begin{IEEEeqnarray}{c}
	\vert j \vert + \vert i-j \vert  \geq  	\vert j + i - j \vert = \vert i \vert
\end{IEEEeqnarray}
As a result, for any $\eta_i (\lambda)$, there is always a stable term of $\lambda^{\vert i \vert}$ that can be factorized in the sum. It is important to note that $\lambda^{\vert i \vert}$ has the same index as  $\eta_i (\lambda)$. Moreover, $\tilde{q}_j(\lambda)$ and $\tilde{v}_{i-j}(\lambda)$ are both stable. Therefore, we can write
\begin{IEEEeqnarray}{rCl}
	\eta_i (\lambda) &=& \lambda^{\vert i \vert} \sum_{j = - \infty}^{\infty} \lambda^{\vert j \vert + \vert i-j \vert -\vert i\vert} \tilde{q}_j(\lambda) \tilde{v}_{i-j}(\lambda) 
				= \lambda^{\vert i \vert} \tilde{\eta}_i (\lambda) \qquad
\end{IEEEeqnarray}
where $\tilde{\eta}_i (\lambda)$ is stable. Substituting (\ref{T_i}) and (\ref{T2outQ}) into the decentralized optimization (\ref{J_simp}) yields
\begin{align}
	J^2 &=& \underset{  \eta_i(\lambda) \ \textmd{is cone stable} }{\text{inf}}
	\big\vert \big\vert \sum_{i = - \infty}^{\infty} \tilde{T}_i (\lambda) z^i -
	\sum_{i = - \infty}^{\infty} \eta_i(\lambda) z^i  \big\vert \big\vert^2_{L_2} \nonumber \\
	&=& \underset{ \tilde{\eta}_i (\lambda) \ \textmd{is stable} }{\text{inf}}
	\big\vert \big\vert \sum_{i = - \infty}^{\infty} \tilde{T}_i (\lambda) z^i -
	\sum_{i = - \infty}^{\infty} \lambda^{\vert i \vert} \tilde{\eta}_i (\lambda) z^i  \big\vert \big\vert^2_{L_2} 
\end{align}
Using the Parseval's identity
\begin{IEEEeqnarray}{c}
	J^2 = \underset{ \tilde{\eta}_i  (\lambda) \ \textmd{is stable} }{\text{inf}}
	\sum_{i = - \infty}^{\infty} \big\vert \big\vert \tilde{T}_i (\lambda) -
	\lambda^{\vert i \vert} \tilde{\eta}_i (\lambda) \big\vert \big\vert^2_{L_2} \label{Pars}
\end{IEEEeqnarray}
where the following equation gives the optimal $\hh$ decentralized cost ($J_{opt}$) for this class of spatially invariant systems
\begin{IEEEeqnarray}{c}	
	J_{opt}^2 =\sum_{i = - \infty}^{\infty} \big\vert \big\vert \frac{\tilde{T}_i (\lambda)}{\lambda^{\vert i \vert}} \big\vert \big\vert^2_{\hh^\perp}
\end{IEEEeqnarray}
The minimum in (\ref{Pars}) is achieved by choosing $\tilde{\eta}_i (\lambda)$ satisfying
	\begin{eqnarray}
		\tilde{\eta}_i (\lambda)= \Pi \left[ \frac{\tilde{T}_i (\lambda)}{\lambda^{\vert i \vert}} \right] \quad
		  \label{cdc45}
	\end{eqnarray}
where $i \in (-\infty, \infty)$ and $\Pi$ is the orthogonal projection from $L_2$ into $\hh$. The optimal decentralized Youla parameter is then
\begin{eqnarray}
	Q = \frac{ \sum\limits_{i=-\infty}^{\infty} \lambda^{\vert i \vert} \tilde{\eta}_i(\lambda)z^i} {T_{2out}(z,\lambda)} \label{Q}
\end{eqnarray}
From (\ref{cdc28}), the explicit optimal $\hh$ decentralized controller $K(z,\lambda)$ is given in the following closed form
\begin{eqnarray}
	K \!=\! - \frac{ \sum\limits_{i=-\infty}^{\infty} \! \lambda^{\vert i \vert} \tilde{\eta}_i(\lambda)z^i}
{T_{2out}(z,\lambda)}    \!\!
    \left[I-G_{yu}(z,\lambda ) \frac{ \sum\limits_{i=-\infty}^{\infty} \! \lambda^{\vert i \vert} \tilde{\eta}_i(\lambda)z^i} {T_{2out}(z,\lambda)}\right]^{-1} \!\!\! \!\!\! \label{cdc48}
\end{eqnarray}
\indent To realize the controller $K(z,\lambda)$, the most straightforward way is to first realize each element of the transfer functions $\sum \lambda^{\vert i \vert}\tilde{\eta}_i(\lambda)z^i$, $T_{2out}$ and $G_{yu}$ individually. Each realization can be obtained by sum or product of several simply realizable transfer function. 
Finally, the optimal $H_2$ decentralized control law $K$ in (\ref{cdc48}) can be combined using basic operations (\ref{SS1})-(\ref{SS3}) and can be realized by a positive feedback interconnection as follows
\begin{figure*}[b] \small \vspace{-0.3cm} \setcounter{equation}{41}
	\hrulefill
	\begin{eqnarray}
	A_k(z) = \left[ \begin{array}{cc} 
	A_3(z) + B_3 D (I-D_3D)^{-1} C_3(z) & B_3 (I-DD_3)^{-1} C(z)  \\ 
	B (I-D_3D)^{-1} C_3(z) & A(z)+BD_3(I-DD_3)^{-1}C(z)
	\end{array} \right], \quad
	B_k = \left[ \begin{array}{c}  -B_3(I-DD_3)^{-1}  \\  -BD_3 (I-DD_3)^{-1}  \end{array} \right] \nonumber\\
	C_k(z)=\left[ \begin{array}{cc}  (I-DD_3)^{-1}C_3(z) & (I-DD_3)^{-1} D_3 C(z) \end{array} \right], \quad D_k=-D_3 (I-D_3D)^{-1} \label{ssK}  \qquad \qquad \qquad
	\end{eqnarray}   \setcounter{equation}{35}
\end{figure*}
\begin{eqnarray} 
	G_{1}(z,\lambda) := \sum\limits_{i=-\infty}^{\infty} \! \lambda^{\vert i \vert} \tilde{\eta}_i(\lambda)z^i =\left[ \begin{array}{c|c}  A_1(z) & B_1 \\ \hline  C_1(z) & D_1  \end{array} \right]\\
	G_{2}(z,\lambda) := T_{2out} (z,\lambda) =\left[ \begin{array}{c|c}  A_2(z) & B_2 \\ \hline  C_2(z) & D_2 \end{array} \right] \quad \\
	G_{yu}(z,\lambda)=\left[ \begin{array}{c|c}  A(z)  & B \\ \hline  C(z)& D  \end{array} \right]  \quad \qquad
\end{eqnarray} 
and we have
\begin{eqnarray}
G_{2}(z,\lambda)^{-1} =\left[ \begin{array}{c|c}  A_2(z)-B_2 D_2^{-1}C_2(z) & -B_2 D_2^{-1} \\ \hline  D_2^{-1}C_2(z) & D_2^{-1}  \end{array} \right] \\
\overbrace{\color{white} \_\_\_\_\_\_\_\_\_\_\_\_\_\_\_\_\_\_\_\_}^{A_3} \quad\qquad \overbrace{\color{white} \_\_\_\_\_}^{B_3} \nonumber\ \quad\\
\frac{G_1(z,\lambda)}{G_2(z,\lambda)}\!=\! \!\left[ \begin{array}{cc|c} \! \! A_1(z)  &\!\! B_1 D_{2}^{-1} C_2(z) \!\!&\!\! B_1 D_{2}^{-1} \!\!\\  
\!\! 0 &\!\! A_2(z)-B_2 D_{2}^{-1}C_2(z) \!\!&\!\! -B_2 D_2^{-1} \!\!\\ \hline
\!\! C_1(z) &\!\! D_1 D_2^{-1} C_2(z) \!\!&\!\! D_1 D_{2}^{-1} \!\! \end{array} \right]\\
\underbrace{\color{white} \_\_\_\_\_\_\_\_\_\_\_\_\_\_\_\_\_\_\_\_}_{C_3} \quad\qquad \underbrace{\color{white} \_\_\_\_\_}_{D_3} \nonumber \ \ \quad \nonumber
\end{eqnarray}
From Fig. \ref{fig:3}, it follows that a state space realization of $K$ is given by
\begin{eqnarray}
K{(z,\lambda)}= \left[ \begin{array}{c|c} A_k(z) & B_k\\ \hline
C_k(z) & D_k \end{array} \right]
\end{eqnarray}
\noindent where $A_k$, $B_k$, $C_k$ and $D_k$ are defined in (42).
\section{Numerical Results}
In this section, the above framework is applied to design an optimal $H_2$ decentralized controller for a numerical example. For comparison purposes, we followed the discrete time example given in \cite{voulgaris2003optimal} obtained by discretizing a specific partial differential equation. The goal is to compute the optimal $H_2$ disturbance attenuation for the system with transfer function $G(z,\lambda)$ and the weighting function $W(z,\lambda)$ given as follows
\begin{eqnarray}\setcounter{equation}{43}
	G(z,\lambda)=\frac{\tau \lambda}{1-(\gamma/2)(z^{-1} + 2\alpha + z)\lambda}  \\
	W(z,\lambda)=\frac{\lambda}{1-(c/2)(z^{-1} + 2a + z)\lambda}
\end{eqnarray}
The weighting function has similar structure as the plant $G(z,\lambda)$. Assume that $\tau=1$, $\gamma=1/3$, $\alpha=1$, $c=1/4$ and $a=1$. The problem set-up
\begin{eqnarray}
	J=\inf_{Q \, {\rm stable} \; {\rm s.t.} (\ref{Eq:Q})}  \ab T_{zw} \ab_\hh
\end{eqnarray}
where
\begin{eqnarray}
	\ab T_{zw} \ab_\hh = \ab (1-GQ)W \ab_\hh  =  \ab T_1-T_2 Q \ab_\hh
\end{eqnarray}
The transfer function $T_1(z,\lambda)$ and $T_2(z,\lambda)$ are as follows
\begin{eqnarray}
	T_1(z,\lambda) &=& \frac{\lambda}{1-r(z)\lambda} \\
	T_2(z,\lambda) &=& \frac{ \tau \lambda^2}{(1-\rho(z)\lambda)(1-r(z)\lambda)}
\end{eqnarray}
and
\begin{eqnarray}
	\rho(z) &=&  z/6 + 1/3 + z^{-1}/6   \\
	r(z) &=&     z/8 + 1/4 + z^{-1}/8
\end{eqnarray}
The following inner-outer factorization is computed as
\begin{eqnarray}
	T_{2in}(z,\lambda) &=& \lambda^2  \\
	T_{2out}(z,\lambda) &=& \frac{\tau}{(1-\rho(z)\lambda)(1-r(z)\lambda) }
\end{eqnarray}
where $T_{2in}$ is an isometry and  $T_{2out}$ is causally invertible with respect to the temporal variable.
It can be seen that
\begin{eqnarray}
	T^{\ast}_{2in}(z,\lambda) T_1(z,\lambda) &=& \frac{\lambda^{-1}}{1-r(z)\lambda} = \lambda^{-1} + r(z) \nonumber\\
	 &+& \lambda r^2(z) + \lambda^2 r^3(z) +  \dots \label{T_i_Ex}
\end{eqnarray}
Using (\ref{T_i}) and (\ref{T_i_Ex}), $\tilde{T}_i (\lambda)$ can be approximated as
\begin{IEEEeqnarray}{c}
	\tilde{T}_0 (\lambda) = \lambda^{-1} + \frac{1}{4} + \frac{3}{32}\lambda^{1} + \frac{5}{128}\lambda^{2} + \nonumber \dots \\
	\tilde{T}_{\pm 1} (\lambda) = \frac{1}{8} + \frac{1}{16}\lambda + \frac{15}{512} \lambda^2 + \frac{7}{512} \lambda^3 + \dots 
\end{IEEEeqnarray}
Therefore, $ \tilde{\eta}_i $ can be calculated as
\begin{IEEEeqnarray}{c}
	\tilde{\eta}_0 (\lambda) = \frac{1}{4} + \frac{3}{32}\lambda^{1} + \frac{5}{128}\lambda^{2} +  \nonumber \dots \\
	\tilde{\eta}_{\pm 1} (\lambda) = \frac{1}{16} + \frac{15}{512} \lambda^1 + \frac{7}{512} \lambda^2 + \dots
\end{IEEEeqnarray}
Note that the problem is infinite dimensional, and we have presented in the above calculations five spatial order. From (\ref{Q}), the Youla parameter $Q$ is then calculated as
\begin{IEEEeqnarray}{rlc}
	 Q(z,\lambda) &=& \frac{1}{4} - \frac{1}{96}(z^{+1} + 5 + z^{-1})\lambda - \frac{1}{1536} (2 z^{+2} \\&&+ 21 z^{+1} + 32 + 21 z^{-1} + 2 z^{-2})\lambda^2 + \dots \nonumber
\end{IEEEeqnarray}
In general, the transfer function $\sum \tilde{\eta}_i(\lambda)z^i$ has an infinite number of terms. As a result, $Q(z,\lambda)$ is also infinite dimensional. By computing the transfer function $Q(z,\lambda)$ for three terms, the distributed controller $K(z,\lambda)$ can be calculated as shown in (\ref{K}). The state-space description of $K(z,\lambda)$ can also be obtained using the procedure in Section III as shown in (\ref{SS_Ex}).\\ \indent
\begin{figure*}[b] \small 
	\hrulefill
	\begin{IEEEeqnarray}{c}
		K \!=\!  \frac{(\!-\! 2z^3 \!-\! 11z^2 \!-\! 26z^1 \!-\! 34 \!-\! 26z^{-1} \!-\! 11z^{-2} \!-\! 2z^{-3})\lambda^3 \!+\! (20z^2 \!+\! 66z^1 \!+\! 92 \!+\! 66z^{-1} \!+\! 20z^{-2})\lambda^2 \!+\! (16z^1 \!+\! 80 \!+\! 16z^{-1})\lambda \!-\! 384}{(12z^2 + 42z^1 + 60 + 42z^{-1} + 12z^{-2})\lambda^3 + (- 48z^1 - 48 - 48z^{-1})\lambda^2 -384\lambda + 1536} \label{K} 
	\end{IEEEeqnarray} 
\small \textbf{\hrulefill}
\begin{equation}
\begin{aligned}
T_{out}^{-1}(z,\lambda) &:
A (z) \!=\! \left[
\begin{array}{cc}
0 & - 0.125 - 0.25 - 0.125 z^{-1}  \\
0 &   0
\end{array} \right] \!,
B \!=\!	\left[
\begin{array}{c}
-1.0  \\
-1.0
\end{array} \right] \!,
C^T(z) \!=\! \left[
\begin{array}{c}
0.167 z + 0.333 + 0.167z^{-1} \\
0.125 z +  0.250 + 0.125z^{-1} 	
\end{array} \right] \!,
D \!=\!	1.0 \\
\sum\limits_{i}\lambda^{\vert i \vert}\tilde{\eta}_i(\lambda)z^i \approx &:
A (z) = 0 , \qquad
B =	1.0,  \qquad
C^T(z) = 0.0625 z + 0.0938 + 0.0625z^{-1}, \qquad
D=0.25 \\
G(z,\lambda) & :
A (z) = 0.167z + 0.333 + 0.167z^{-1}, \qquad
B =	1.0,  \qquad
C^T(z) = 1.0, \qquad
D=0 \\
K(z,\lambda) &:
A (z) \!=\! \left[
\begin{array}{ccc}
0& - 0.125 z - 0.25 - 0.125z^{-1} & - 0.0625 z  - 0.0938 - 0.0625z^{-1} \\
0&                          0     & - 0.0625 z  - 0.0938 - 0.0625z^{-1}\\
0&                          0     &                              0     \\
-0.167 z - 0.333 - 0.167z^{-1} 	& - 0.125 z - 0.25 - 0.125z^{-1} & - 0.0625 z  - 0.0938 - 0.0625z^{-1}\\
\end{array} \right.\quad \\
& \qquad   \left.	\begin{array}{c}
0.25\\
0.25\\
-1\\
0.167 z + 0.583 + 0.167z^{-1}
\end{array} \right],
B \!=\!	\left[
\begin{array}{c}
-0.25\\
-0.25\\
1.0\\
-0.25
\end{array} \right],
C^T(z) \!=\! \left[
\begin{array}{c}
- 0.167 z - 0.333 - 0.167z^{-1} \\
- 0.125 z -  0.250 - 0.125z^{-1} \\
- 0.0625 z - 0.0938 - 0.0625z^{-1} \\
0.25
\end{array} \right] \!,
D = 0.25 \quad \end{aligned} \label{SS_Ex}
\end{equation}
\end{figure*}
Table \ref{Results} shows the resulting closed-loop performance for the optimal decentralized controller with different approximation order as well as other types of decentralized controllers. It is interesting to note that our method converges very fast to the optimal decentralized norm. In \cite{voulgaris2003optimal}, the authors had calculated the solution of the relaxed controller and the optimal decentralized norm numerically. There was no explicit solution on the non-relaxed decentralized controller $K$ (or Youla parameter $Q$). It can clearly be seen that our proposed method achieves better performance in comparison to the relaxed controller.
\begin{table}[]
	\renewcommand{\arraystretch}{1}
	\centering
	\caption{Comparison of Optimal Norm for Different Temporal Orders of the Distributed Controller.}
	\label{Results}
	\begin{tabular}{|c|c|c|}
		\hline
		\multicolumn{2}{|c|}{Temporal Order} & \multirow{2}{*}{$J=\ab T_{zw} \ab_\hh$} \\ \cline{1-2}
		$\sum \tilde{\eta}_i(\lambda)z^i$  &  $Q(z,\lambda) $  &   \\ \hline
			0	&      2      &     1.0261    \\ \hline
		    1	&      3      &     1.0180    \\ \hline
		    2	&      4      &     1.0162    \\ \hline
		    3	&      5      &     1.0159    \\ \hline
		    4	&      6      &     1.0158    \\ \hline
		    5	&      7      &     1.0158    \\ \hline
		    6	&      8      &     1.0157    \\ \hline\hline
		    \multicolumn{2}{|c|}{Using Relaxed Controller in \cite{voulgaris2003optimal}}   &  1.0659   \\ \hline
		    \multicolumn{2}{|c|}{Optimal Decentralized Norm \cite{voulgaris2003optimal}}       &     1.0157    \\ \hline
		    \multicolumn{2}{|c|}{Using Centralized Controller \cite{voulgaris2003optimal}}      			 &     1.0000       \\ \hline
	\end{tabular}
\end{table} 
\section{Conclusion}
In this work, we have developed a method to design the optimal $\hh$ decentralized controller for a class of spatially invariant systems. The decentralized controller assumed same structure as the plant whose impulse response admits a cone structure. Using Parseval's identity, the optimal $\hh$ decentralized control problem is transformed into an infinite number of model matching problems with a specific structure that can be solved efficiently. In addition, the closed-form expression (explicit formula) of the decentralized controller is derived for the first time. Moreover, a constructive procedure to obtain the state-space representation of the decentralized controller which is more convenient for implementation. An illustrative numerical example is presented. In a forthcoming paper, the control design of optimal $\hh$ decentralized control laws for funnel causal spatially invariant systems will be studied. 
{\small
	\bibliographystyle{IEEEtran}
	\bibliography{Ref_Distributed}

\begin{thebibliography}{10}
\providecommand{\url}[1]{#1}
\csname url@samestyle\endcsname
\providecommand{\newblock}{\relax}
\providecommand{\bibinfo}[2]{#2}
\providecommand{\BIBentrySTDinterwordspacing}{\spaceskip=0pt\relax}
\providecommand{\BIBentryALTinterwordstretchfactor}{4}
\providecommand{\BIBentryALTinterwordspacing}{\spaceskip=\fontdimen2\font plus
\BIBentryALTinterwordstretchfactor\fontdimen3\font minus
  \fontdimen4\font\relax}
\providecommand{\BIBforeignlanguage}[2]{{%
\expandafter\ifx\csname l@#1\endcsname\relax
\typeout{** WARNING: IEEEtran.bst: No hyphenation pattern has been}%
\typeout{** loaded for the language `#1'. Using the pattern for}%
\typeout{** the default language instead.}%
\else
\language=\csname l@#1\endcsname
\fi
#2}}
\providecommand{\BIBdecl}{\relax}
\BIBdecl

\bibitem{mahajan2012information}
A.~Mahajan, N.~C. Martins, M.~C. Rotkowitz, and S.~Y{\"u}ksel, ``Information
  structures in optimal decentralized control,'' in \emph{IEEE Conf. on
  Decision and Control (CDC)}, 2012, pp. 1291--1306.

\bibitem{mahajan2008identifying}
A.~Mahajan, A.~Nayyar, and D.~Teneketzis, ``Identifying tractable decentralized
  control problems on the basis of information structure,'' in \emph{Conf. on
  Comm., Control, and Computing}, 2008, pp. 1440--1449.

\bibitem{mahajan2009optimal}
A.~Mahajan and D.~Teneketzis, ``Optimal performance of networked control
  systems with nonclassical information structures,'' \emph{SIAM Journal on
  Control and Optimization}, vol.~48, no.~3, pp. 1377--1404, 2009.

\bibitem{nayyar2011optimal}
A.~Nayyar, A.~Mahajan, and D.~Teneketzis, ``Optimal control strategies in
  delayed sharing information structures,'' \emph{IEEE Trans. on Automatic
  Control}, vol.~56, no.~7, pp. 1606--1620, 2011.

\bibitem{Ehsan}
M.~E. Raoufat and S.~M. Djouadi, ``Control allocation for wide area coordinated
  damping,'' in \emph{IEEE PES General Meeting}, 2017, pp. 1--5.

\bibitem{wolfe1996decentralized}
J.~Wolfe, D.~Chichka, and J.~Speyer, ``Decentralized controllers for unmanned
  aerial vehicle formation flight,'' in \emph{Guidance, Navigation, and Control
  Conf.}, 1996, p. 3833.

\bibitem{bamieh1999optimal}
B.~Bamieh, F.~Paganini, and M.~Dahleh, ``Optimal control of distributed arrays
  with spatial invariance,'' in \emph{Robustness in identification and
  control}.\hskip 1em plus 0.5em minus 0.4em\relax Springer, 1999, pp.
  329--343.

\bibitem{barooah2007control}
P.~Barooah, P.~G. Mehta, and J.~P. Hespanha, ``Control of large vehicular
  platoons: Improving closed loop stability by mistuning,'' in \emph{American
  Control Conf. (ACC)}, 2007, pp. 4666--4671.

\bibitem{radner1962team}
R.~Radner, ``Team decision problems,'' \emph{The Annals of Mathematical
  Statistics}, vol.~33, no.~3, pp. 857--881, 1962.

\bibitem{witsenhausen1968counterexample}
H.~S. Witsenhausen, ``A counterexample in stochastic optimum control,''
  \emph{SIAM Journal on Control}, vol.~6, no.~1, pp. 131--147, 1968.

\bibitem{ho1972team}
Y.-C. Ho \emph{et~al.}, ``Team decision theory and information structures in
  optimal control problems--part i,'' \emph{IEEE Trans. on Automatic control},
  vol.~17, no.~1, pp. 15--22, 1972.

\bibitem{bamieh2002distributed}
B.~Bamieh, F.~Paganini, and M.~A. Dahleh, ``Distributed control of spatially
  invariant systems,'' \emph{IEEE Trans. on Automatic Control}, vol.~47, no.~7,
  pp. 1091--1107, 2002.

\bibitem{rotkowitz2010parametrization}
M.~C. Rotkowitz, ``Parametrization of all stabilizing controllers subject to
  any structural constraint,'' in \emph{IEEE Conf. on Decision and Control
  (CDC)}, 2010, pp. 108--113.

\bibitem{rotkowitz2006characterization}
M.~Rotkowitz and S.~Lall, ``A characterization of convex problems in
  decentralized control,'' \emph{IEEE Trans. on Automatic Control}, vol.~51,
  no.~2, pp. 274--286, 2006.

\bibitem{lamperski2013output}
A.~Lamperski and J.~C. Doyle, ``Output feedback h2 model matching for
  decentralized systems with delays,'' in \emph{American Control Conf. (ACC)},
  2013, pp. 5778--5783.

\bibitem{kim2015explicit}
J.-H. Kim and S.~Lall, ``Explicit solutions to separable problems in optimal
  cooperative control,'' \emph{IEEE Trans. on Automatic Control}, vol.~60,
  no.~5, pp. 1304--1319, 2015.

\bibitem{voulgaris2003optimal}
P.~G. Voulgaris, G.~Bianchini, and B.~Bamieh, ``Optimal h2 controllers for
  spatially invariant systems with delayed communication requirements,''
  \emph{Systems \& Control Letters}, vol.~50, no.~5, pp. 347--361, 2003.

\bibitem{bamieh2005convex}
B.~Bamieh and P.~G. Voulgaris, ``A convex characterization of distributed
  control problems in spatially invariant systems with communication
  constraints,'' \emph{Systems \& Control Letters}, vol.~54, pp. 575--583,
  2005.

\bibitem{djouadi2014duality}
S.~M. Djouadi and J.~Dong, ``Duality of the optimal distributed control for
  spatially invariant systems,'' in \emph{American Control Conf. (ACC)}, 2014,
  pp. 2214--2219.

\bibitem{djouadi2015operator}
S.~M. Djouadi and J.~. Dong, ``Operator theoretic approach to the optimal
  distributed control problem for spatially invariant systems,'' in
  \emph{American Control Conf. (ACC)}, 2015, pp. 2613--2618.

\bibitem{djouadi2015distributed}
S.~M. Djouadi and J.~Dong, ``On the distributed control of spatially invariant
  systems,'' in \emph{IEEE Conf. on Decision and Control (CDC)}, 2015, pp.
  549--554.

\bibitem{duren2000theory}
P.~Duren, ``Theory of hp spaces. mineola,'' 2000.

\bibitem{fardad2011design}
M.~Fardad and M.~R. Jovanovi{\'c}, ``Design of optimal controllers for
  spatially invariant systems with finite communication speed,''
  \emph{Automatica}, vol.~47, no.~5, pp. 880--889, 2011.

\bibitem{zhou1998essentials}
K.~Zhou and J.~C. Doyle, \emph{Essentials of robust control}.\hskip 1em plus
  0.5em minus 0.4em\relax Prentice hall Upper Saddle River, NJ, 1998, vol. 104.

\bibitem{vidyasagar2011control}
M.~Vidyasagar, ``Control system synthesis: a factorization approach, part ii,''
  \emph{Synthesis lectures on control and mechatronics}, vol.~2, no.~1, pp.
  1--227, 2011.

\end{thebibliography}
}
\end{document}